\documentclass[doublecol]{epl2} 

\usepackage{float}
\usepackage{bm}
\usepackage{array}
\usepackage{amsmath}
\usepackage{booktabs} 
\usepackage{multicol}

\graphicspath{{Pictures/}}

\title{Correlation and scaling behaviors of $PM_{2.5}$ concentration in China}
\shorttitle{Correlation and scaling behaviors of $PM_{2.5}$ concentration in China} 

\author{Yongwen. Zhang\inst{1,2} \and Dean. Chen\inst{3} \and Jingfang. Fan\inst{4} \and Shlomo. Havlin\inst{4} \and Xiaosong. Chen\inst{2,5} \thanks{E-mail: \email{chenxs@itp.ac.cn}}}
\shortauthor{Yongwen. Zhang  \etal}

\institute{                    
  \inst{1} Data Science Research Center, Kunming University of Science and Technology, Kunming 650500, Yunnan, China\\
  \inst{2} CAS Key Laboratory of Theoretical Physics, Institute of Theoretical Physics, Chinese Academy of Sciences, P. O. Box 2735, Beijing 100190, China\\
  \inst{3} Department of Physics, University of Helsinki, P.O. Box 48, 00014 Helsinki, Finland\\
  \inst{4} Department of Physics, Bar-Ilan University, Ramat-Gan 52900, Israel\\
  \inst{5} School of Physical Sciences, University of Chinese Academy of Sciences, Beijing 100049, China;  }


\abstract{
Air pollution has become a major issue and caused widespread environmental and health problems. Aerosols or particulate matters are an important component of the atmosphere and can transport under complex meteorological conditions. Based on the data of $PM_{2.5}$ observations, 
we develop a network approach to study and quantify their spreading and diffusion patterns. We calculate cross-correlation functions of time lag between sites within different season.  The probability distribution of correlation changes with season. It is found that the probability distributions in four seasons can be scaled into one scaling function with averages and standard deviations of correlation. This seasonal scaling behavior indicates there is the same mechanism behind correlations of $PM_{2.5}$ concentration in different seasons. 
Further, from weighted and directional degrees of complex network, different properties of $PM_{2.5}$ concentration are studied. The weighted degrees reveal the strongest correlations of $PM_{2.5}$ concentration in winter and in the North China plain. These directional degrees show net influences of $PM_{2.5}$ along Gobi and inner Mongolia, the North China plain, Central China, and Yangtze River Delta.}

\begin{document}

\maketitle

\section{Introduction}

Aerosols or particulate matters, which control process from low visibility events to precipitation, are important components of the atmosphere. They play a critical role in global climate pattern and public health. Chen et al. \cite{chen2013evidence} have reported the impact on life expectancy of sustained exposure to air pollution from China 's Huai River policy. Due to anthropogenic emissions, the concentration of particulate matters is growing sharply. In the past few years, China has witnessed rapid growth both in industry and in city’s population. As a result, air pollution, especially the pollution caused by high $PM_{2.5}$ concentration, has become a serious issue \cite{Cao}. 

Most previous studies on $PM_{2.5}$ concentrated on observation in one site. Winter and summer $PM_{2.5}$ chemical compositions in 14 cities of China have been analysed by Cao et al. \cite{Cao2012}. The publishing of hourly data since 2013 provided possibility to study spatial distribution and seasonal variation of $PM_{2.5}$ in China \cite{YZhang}. Using the data of monitoring network in the North China Plain and the Yangtze River Delta, Hu et al. \cite{Hu} found strong temporal correlation between cities within $250$ km. For $81$ cities in China, Gao et al. \cite{Gao} studied air pollution of city clusters from June 2004 to June 2007.  The relation between air quality  over Beijing and its surroundings and circulation patterns was studied by Zhang et al. \cite{JZhang}.  The spatiotemporal variations of $PM_{2.5}$ and $PM_{10}$ concentrations of 31 Chinese cities from March 2013 to March 2014 were related to SO$_2$, NO$_2$, CO and O$_3$ \cite{Xie}. At a suburban site between Beijing and Tianjin, the correlation of pollutants to meteorological conditions was discussed \cite{Xu}.

The studies \cite{Hu,Gao} have shown that $PM_{2.5}$ concentrations in different cities are not localized and related each other. It is of great interest to investigate how far the $PM_{2.5}$ concentrations in different cities of China are correlated. Using the hourly data of monitoring sites over China, the spatial correlations of $PM_{2.5}$ concentrations in 2015 have been studied using the principal component analysis \cite{Di}. 

In the last decade, network has emerged as an important tool in studies of complex systems and has been applied to a wide variety of disciplines \cite{Boccaletti,Newman2002,RCohen}.  Recently, complex network theory has been used to study climate systems \cite{wiedermann,Boers,Ludescher,Tsonis,Yamasaki2008,YWang2013,YWang2016,Berezin}.
For a climate system, geographical locations or grid points are regarded as nodes of network and the links between them are defined from a cross correlation function \cite{Yamasaki2008,Berezin} or event synchronization \cite{JDong}. 

In this letter, we study the $PM_{2.5}$ concentrations in China from the aspect of complex networks. The nodes of $PM_{2.5}$ concentration network can be defined from the monitoring stations. Using $PM_{2.5}$ concentration data, we can calculate the correlation between nodes and define their links. The global properties of $PM_{2.5}$ concentrations in China will be studied from the aspect of network. Our work is organized as follows. In the next section, we describe the data and introduce the methodology. The results are presented and discussed in the third section. Finally, a short summary is given.

\section{Data and Methodology}
\label{Data and  Methodology}

\subsection{Data}
The Ministry of Environmental Protection of China has been publishing air quality index since 2013 and  provide data for us to study atmospheric pollution. We use the hourly $PM_{2.5}$ concentration data of $754$ monitoring sites over China from Dec.2014 to Nov.2015 (http://113.108.142.147:20035/emcpublish/). In pre-processing, we transform $754$ monitoring stations into $163$ sites with the 
area $1^{\circ} \times 1^{\circ}$. The concentration of a site is defined by the average of monitoring stations inside this site. Since the strong seasonal dependence of $PM_{2.5}$  concentration, we divide the data into four groups corresponding to winter(Dec, Jan, Feb), spring(Mar, Apr, May), summer(Jun, Jul, Aug) and autumn(Sep, Oct, Nov). 

\subsection{Methodology}
During a time period $T$, the $PM_{2.5}$ concentration of site $i$ has a series $X_{i} (t)$. With respect to its average $\left\langle X_{i} \right\rangle =\frac{1}{T}\sum_{t=1}^{T} X_{i} (t)$, there is a fluctuation series $\delta X_{i} (t)= X_{i} (t)-\left\langle X_{i} \right\rangle$. To study the correlation of $PM_{2.5}$ concentration between sites $i$ and $j$, we calculate the cross-correlation function \cite{Yamasaki2008},
\begin{equation}
\hat C_{ij}(\tau)=\frac{\left\langle \delta X_{i}(t) \cdot \delta X_{j}(t+\tau) \right\rangle}{\sqrt{\left\langle \left[ \delta X_{i} (t) \right]^{2} \right\rangle} \cdot \sqrt{\left\langle \left[ \delta X_{j} (t+\tau)\right]^{2} \right\rangle}}\;,
\label{corr}
\end{equation}
where $-\tau_{max} \le \tau \le \tau_{max}$ is the time lag. On the basis of time-reversal symmetry, there is a relation $\hat C_{ij}(-\tau)=\hat C_{ji}(\tau)$. The cross-correlation in the interval $[-\tau_{max},\tau_{max}]$ can be calculated by $\hat C_{ij}(\tau\geq0)$ and $\hat C_{ji}(\tau\geq0)$. We identify the largest absolute value of $\hat C_{ij}(\tau)$ and denote the corresponding time lag as $\tau_{ij}^*$. The correlation between sites $i$ and $j$ is defined as $C_{ij} \equiv \hat C_{ij}(\tau^*)$. If $\tau_{ij}^* \neq 0$, the correlation between sites $i$ and $j$ is directional. The direction of correlation is from $i$ to $j$ when $\tau_{ij}^* > 0$ and from $j$ to $i$ when $\tau_{ij}^*< 0$.

For given $N$ nodes, there are $(N-1)N/2$ correlations and they can be described by a probability distribution function (PDF) $\rho (C)$. 

For the definition of a network,  a threshold $\Delta$ of correlation is introduced to exclude noise. The adjacency matrix of the network is defined with the threshold as
\begin{equation}
A_{ij}=
\begin{cases}
1-\delta_{ij} &  |C_{ij}| > \Delta \\
0 & |C_{ij}| \leq \Delta\;,
\end{cases}
\label{admatrixA}
\end{equation}
where the Kronecker's delta $\delta_{ij}=0$ for $i \neq j$ and $\delta_{ij}=1$ for $i = j $ so that self-loop is excluded.

The importance of site $i$ in the network is characterized usually by its degree $k_{i}^C=\sum_{j=1}^{N} A_{ij}$ \cite{Boccaletti}. More information can be taken into account with a weighted degree 
\begin{equation}
\bar k_{i}^C=\sum_{j=1}^{N} A_{ij}\left|C_{ij}\right|\;.
\label{Deg}
\end{equation}

The direction from sites $i$ to $j$ is described by a unit vector $\vec {e}_{ij}=\frac{1}{d}\left( \delta \phi, \delta \theta \right)$ with $d = \sqrt{\delta \phi^2 + \delta \theta^2}$, where $\delta \phi$ and $\delta \theta$ are the longitude and latitude differences of $i$ and $j$ respectively. We can further introduce a directional degree as
\begin{equation}
\vec{k}_{i}^C=\sum_{j=1, \tau^{*}_{ij} > 0}^{N}A_{ij}\left|C_{ij}\right|\ \vec{e}_{ij}+\sum_{j=1, \tau^{*}_{ij} < 0}^{N}A_{ij}\left|C_{ij}\right|\ (-\vec{e}_{ij})
\label{Vec}
\end{equation}
to quantify the $PM_{2.5}$ concentration directional influences of site $i$.

Alternatively, we can determine network links according to
\begin{equation}
G_{ij}=\frac{C_{ij}-mean(\hat C_{ij}(\tau))}{std(\hat C_{ij}(\tau))}\;,
\label{G}
\end{equation}
where ``mean'' and ``std'' represent the mean and standard deviation of the cross-correlation function \cite{guez2014influence,YWang2013,fan2017network}. 

The adjacency matrix of network is now defined as 
\begin{equation}
B_{ij}=
\begin{cases}
1-\delta_{ij} &  |G_{ij}| > \Theta \\
0 & |G_{ij}| \leq \Theta\;.
\end{cases}
\label{admatrixB}
\end{equation}
with the threshold $\Theta$ of $G$. With $C_{ij}$ replaced by $G_{ij}$ in Eq. (\ref{Deg}) and Eq. (\ref{Vec}), we can obtain the weighted degree $\bar k_{i}^G$ and the directional degree $\vec{k}_{i}^G$ of $G$.

\section{Results}
\label{RESULTS}

\begin{figure}[H]
\onefigure[width=20pc]{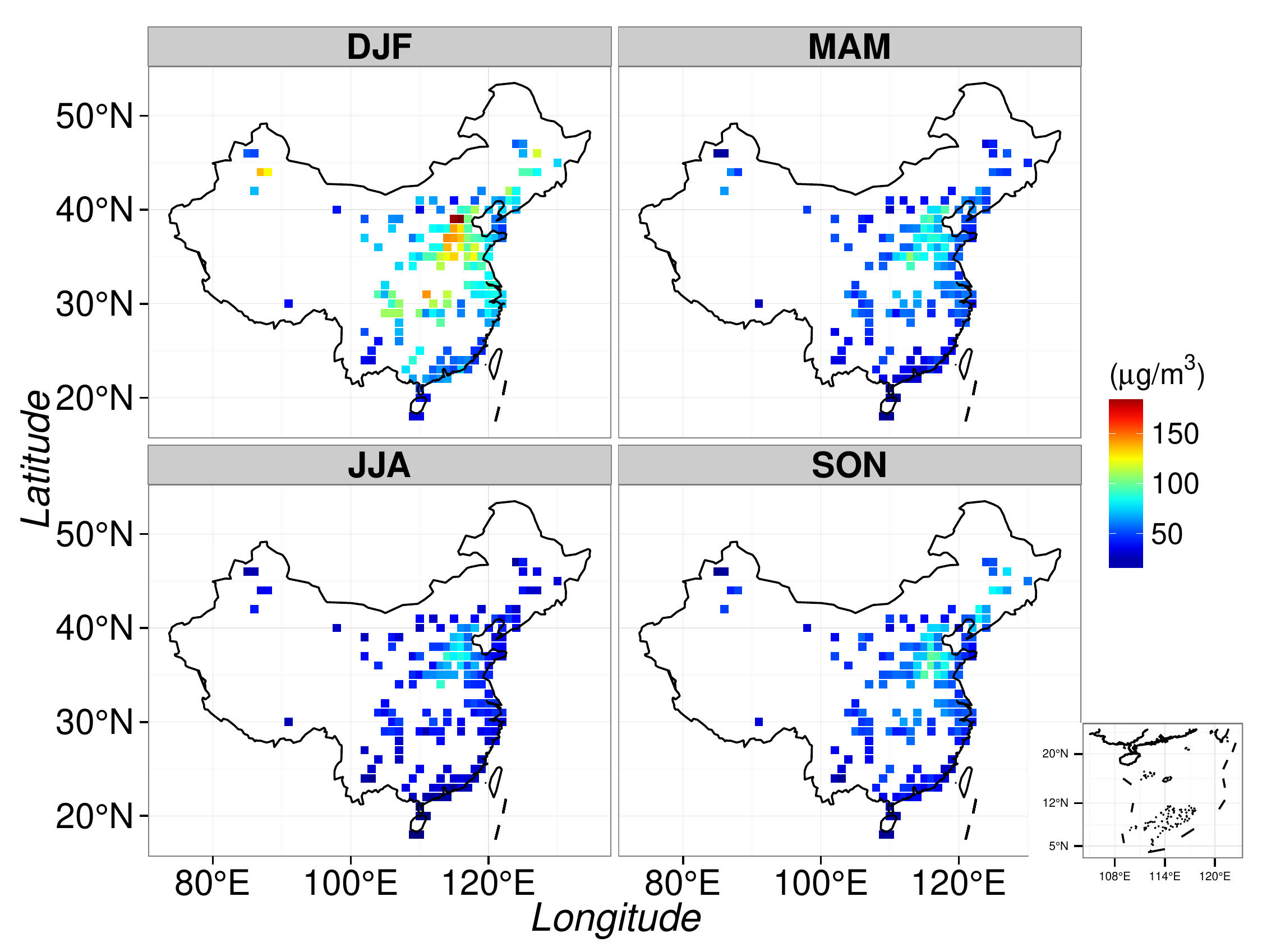}
\centering
\caption{(Color online) Distribution of mean $PM_{2.5}$ concentration over China for the four seasons of 2015.
}
\label{mean}
\end{figure}

We calculate firstly the mean $PM_{2.5}$ concentration $\left\langle X_{i} \right\rangle =\frac{1}{T}\sum_{t=1}^{T} X_{i} (t)$ of sites $i=1,2,...,163$ and show them in Fig. \ref{mean} for four seasons. The overall average 
\begin{equation}
\bar X = \frac{1}{N} \sum_{i=1}^{N} \left\langle X_{i} \right\rangle
\end{equation}
is $75.3 \mu g/m^3$ in winter, $48.2 \mu g/m^3$ in spring, $36.5 \mu g/m^3$ in summer and $46.9 \mu g/m^3$ in autumn. In winter, $45$ percent of sites has mean $PM_{2.5}$ concentration above $75 \mu g/m^3$ and the percentage of the sites above $35 \mu g/m^3$ is $95\%$. The maximum mean $PM_{2.5}$ concentration in winter is related to the enhanced anthropogenic emissions from fossil fuel combustion, biomass burning and unfavorable meteorological conditions for pollution dispersion\cite{YZhang}. In spring,  $8$ percent of sites have mean concentrations above $75 \mu g/m^3$ and $77$ percent are above $35 \mu g/m^3$. The lowest mean $PM_{2.5}$ concentration is reached in summer. Only $4$ percent of sites have mean concentrations above $75 \mu g/m^3$ and $47$ percent are above $35 \mu g/m^3$. In autumn, the percentage of sites above $35\mu g/m^3$ and $75 \mu g/m^3$ reaches $78\%$ and $7\%$, respectively.

\begin{figure}[H]
\onefigure[width=15pc]{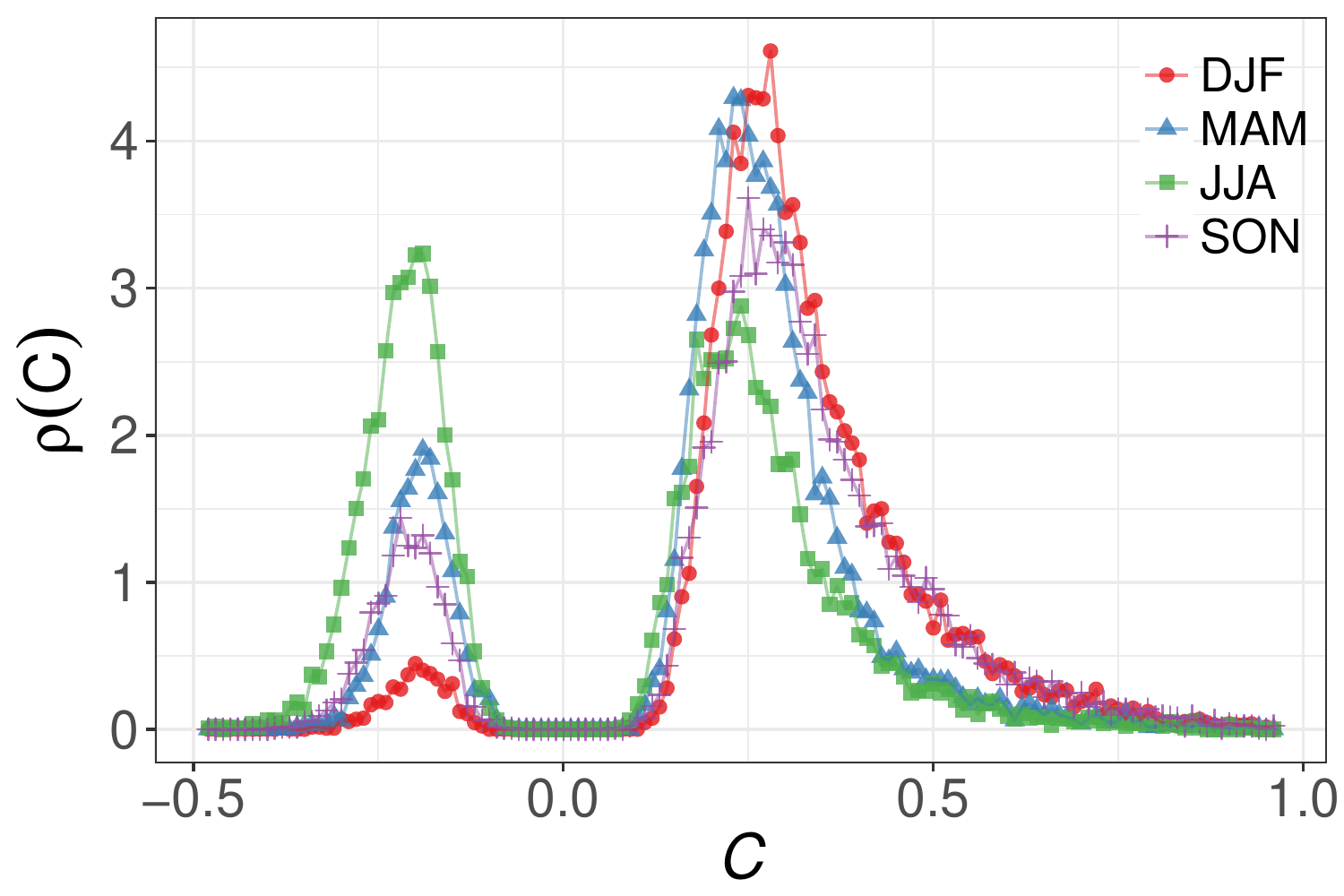}
\centering
\caption{ Probability distribution function of correlation between sites in the four seasons of 2015.
}
\label{num}
\end{figure}

The cross-correlation functions $\hat C_{ij}(\tau)$ between $N$ sites were calculated according to Eq. (\ref{corr}) and with $\tau_{max}=10$ days. From $\hat C_{ij}(\tau)$, we can obtain the correlation $C_{ij}$ between sites $i$ and $j$.  The PDF $\rho (C)$ of correlation is presented in Fig. \ref{num} for four seasons. It can be seen that $\rho (C)$ is separated into positive and negative parts.

\begin{table}[H]
\caption{Proportion, average, and standard deviations of positive and negative correlations.}\label{t1}
\centering
\begin{tabular}{l l l l l l}
\toprule
 & \textbf{DJF} & \textbf{MAM} & \textbf{JJA} & \textbf{SON} \\
\midrule
$\lambda_p$ & $95\%$ & $78\%$ & $55\%$ & $84\%$  \\
$\left\langle C_{p} \right\rangle$ & 0.406 & 0.361 & 0.356 & 0.397 \\
$\sigma_{p}$ & 0.146 & 0.136 & 0.145 & 0.155 \\
\midrule
$\lambda_n$ & $5\%$ & $22\%$ & $45\%$ & $16\%$  \\
$\left\langle C_{n} \right\rangle$ & -0.249 & -0.255 & -0.277 & -0.254 \\
$\sigma_{n}$ & 0.058 & 0.060 & 0.072 & 0.064 \\
\bottomrule
\end{tabular}
\end{table}

\begin{figure}[H]
\onefigure[width=20pc]{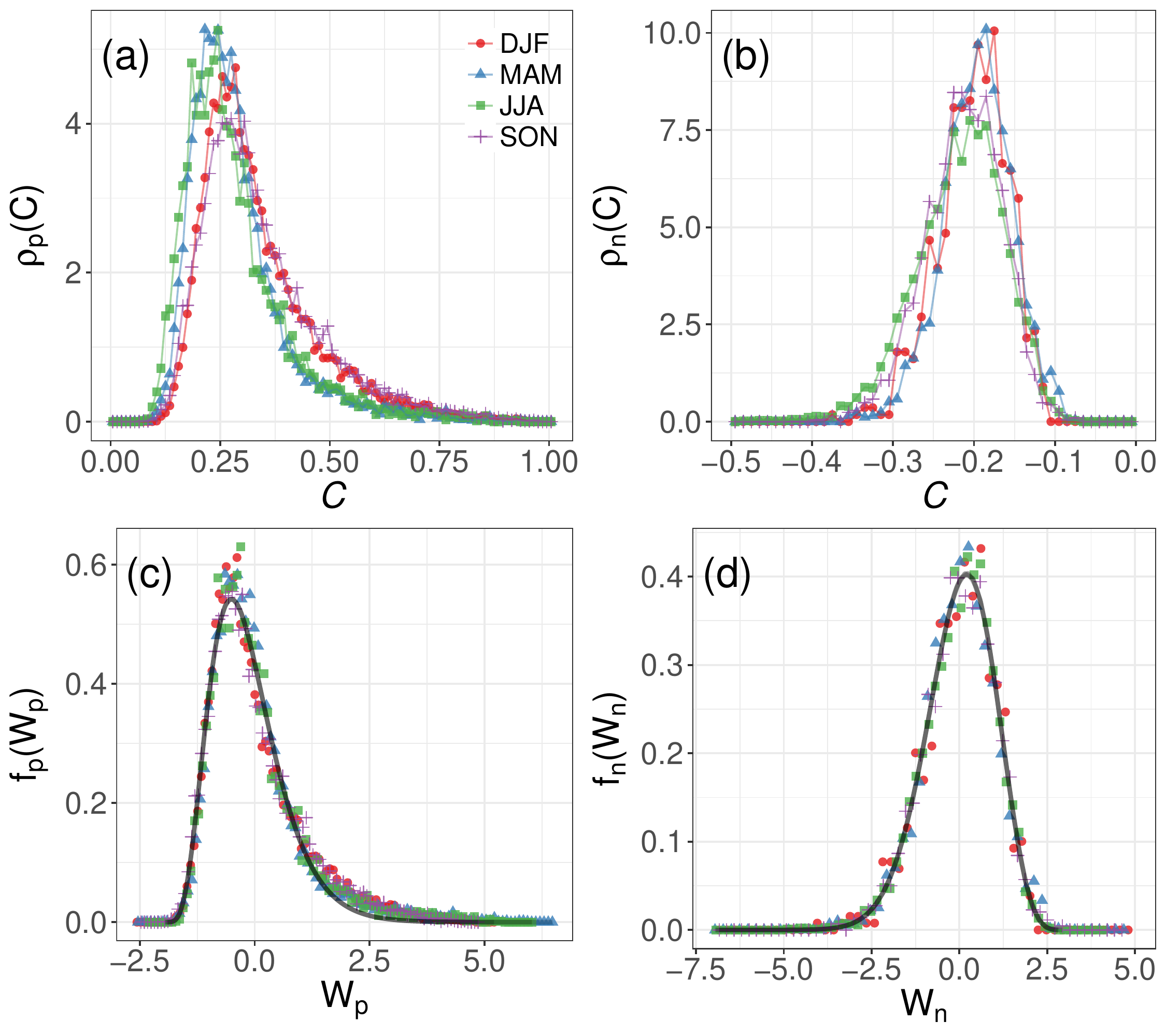}
\centering
\caption{ (Color online) Probability distribution functions $\rho_p (C)$ in (a) and $\rho_n (C)$ in (b).
The variation of $f(W)$ as a function of the scaling quantity $W$ for (c) positive and (d) negative correlations.}\label{scaling}
\end{figure}

The proportions of positive and negative correlations can be calculated by
\begin{eqnarray} 
\lambda_p&=&\int_{0}^{1}  \rho (C) d C \;, \\
\lambda_n &=& \int_{-1}^{0} \rho (C) d C\;.
\end{eqnarray}
For positive correlations, we get $\lambda_p=95\%$ in winter, $78\%$ in spring, $55\%$ in summer and $84\%$ in autumn. Correspondingly, the negative correlations have the proportion $\lambda_n=1-\lambda_p=5\%$, $22\%$, $45\%$ and $16\%$ in the four seasons.  

Further, we introduce probability distribution functions 
\begin{equation}
\rho_p (C)=\frac{1}{\lambda_p} \rho (C)
\end{equation}
for $C > 0$ and
\begin{equation}
\rho_n (C)=\frac{1}{\lambda_n} \rho (C)
\end{equation}
for $C < 0$. They are presented in Fig.\ref{scaling} (a) and (b) and depend on season. The averages $\left\langle C_{p} \right\rangle$, $\left\langle C_{n} \right\rangle$ and standard deviations $\sigma_{p}$, $\sigma_{n}$ of positive and negative correlation can be calculated with $ \rho_p (C) $ and $\rho_n (C)$. Their results are summarized in Table \ref{t1} for different seasons.
$\lambda_p$ and $\left\langle C_{p} \right\rangle$ have their maximum in winter and minimum in summer, which is in accord with the overall average of mean $PM_{2.5}$ concentration.

In a system near its critical point, its physical properties follow scaling behavior because of long-range correlation \cite{IsingScaling,fan2012continuous}. The two-variable function of a physical property can be rewritten as a function of scaled variable, which is universal. We take account of long-range correlation of $PM_{2.5}$ concentration and search for scaling behavior of probability distribution functions $\rho_p (C)$ and $\rho_n (C)$. Using the scaling variable
\begin{eqnarray}
W_p &=& \left[C -\left\langle C_p \right\rangle\right]/\sigma_p\;,  \\
W_n &=& \left[C -\left\langle C_n \right\rangle\right]/\sigma_n\;, 
\end{eqnarray}
we can introduce two scaling functions 
\begin{equation}
f_{p}(W_{p})
=\sigma_{p}\cdot\rho_{p}(C)
\end{equation}
for positive correlations and 
\begin{equation}
f_{n}(W_{n})
=\sigma_{n}\cdot\rho_{n}(C).
\end{equation}
for negative correlations. As shown in Fig.\ref{scaling} (c) and (d), the scaling distribution functions for positive and negative correlations in four seasons collapse together. This indicates that there is the same mechanism behind the correlation of $PM_{2.5}$ concentration. 

\begin{figure}
\onefigure[width=20pc]{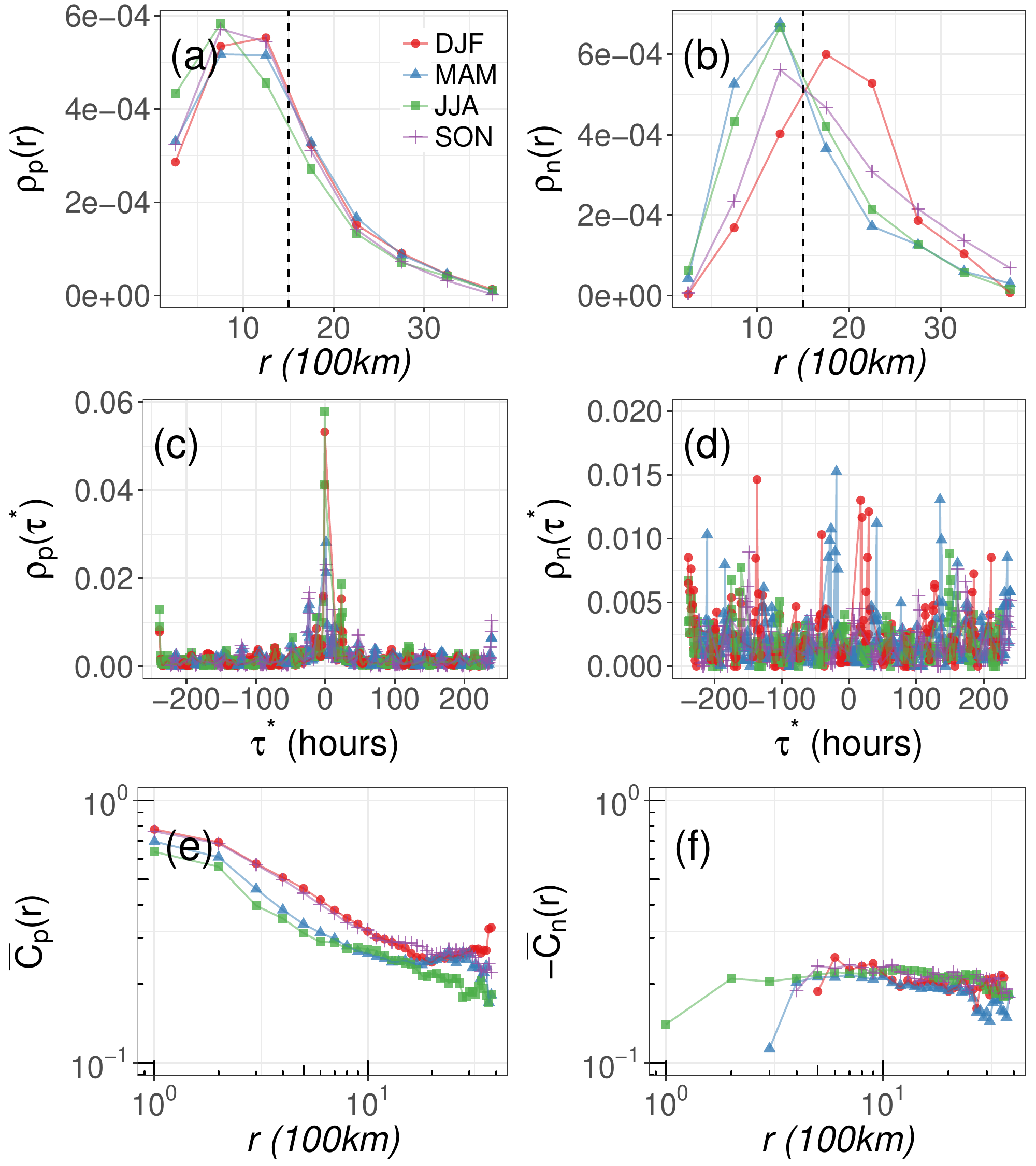}
\centering
\caption{ PDF of distance $r$ is shown in (a) for positive and (b) for negative correlations.       
PDF of time lag $\tau^*$ is shown in (c) for positive and (d) for negative correlations. Averages of positive and negative correlations at distance $r$ are plotted in (e) and (f).}\label{ttdd}
\end{figure}

\begin{figure}
\onefigure[width=15pc]{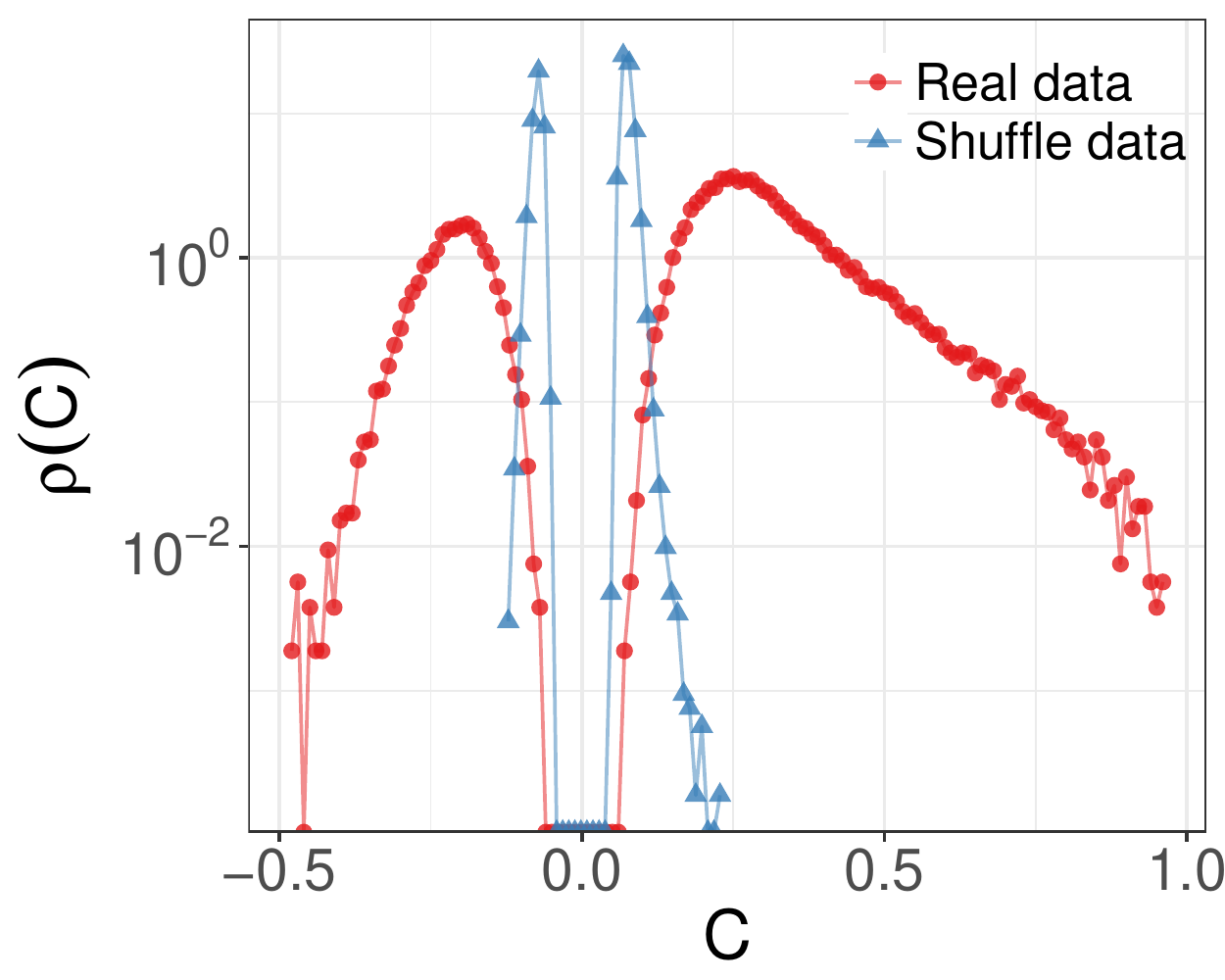}
\centering
\caption{ (Color online) PDF of correlations from real data and shuffle data in all seasons . 
}
\label{p5}
\end{figure}     


The different characters of positive and negative correlations can be demonstrated further by their PDF of distance $r$ and time lag $\tau^*$, which are shown in Fig.\ref{ttdd}.  $\rho_n (r)$ of negative correlations has its peak at a the PDF of $r$ and $\tau^*$ are shown in Fig.\ref{ttdd} (a) and (c). The PDF of negative correlations are presented in Fig.\ref{ttdd} (b) for distance and (d) for time lag. At the peaks of PDF,   the distance of negative correlations is obviously larger than that of positive correlations. The PDF of time lag has maximum at $\tau^*=0$ for positive correlations and $\tau^*\neq 0$ for negative correlations. 
Negative correlations take on the character of larger distance and longer time lag.
 
The average positive correlation $\bar C_p (r)$ at fixed distance $r$ is shown in Fig. \ref{ttdd} (e). In winter and autumn, the decay of $\bar C_p (r)$ follows a power law in some range of $r$.  This could be related to the transport of $PM_{2.5}$ by atmospheric currents. This trend will be weakened in spring and summer\cite{Lau}. The average negative correlation $\bar C_n (r)$ demonstrates quite different behaviors, which are shown in Fig. \ref{ttdd} (f). At large distance, $\bar C_n (r)$ becomes nearly constant. We suppose that negative correlations are resulted by some external factors existing in large scale of distance. 

To define the network of correlation, the threshold $\Delta$ of correlation is determined from the shuffled data obtained by permuting randomly the real data in a season.  PDF of correlation from shuffle data is compared with that from real data in Fig. \ref{p5}. 
We define the average of absolute values of correlations from shuffled data as the threshold $\Delta$. We obtain $\Delta=0.017$ and the adjacency matrix of the network for correlation $C$ according to Eq. (\ref{admatrixA}).

\begin{figure}
\onefigure[width=20pc]{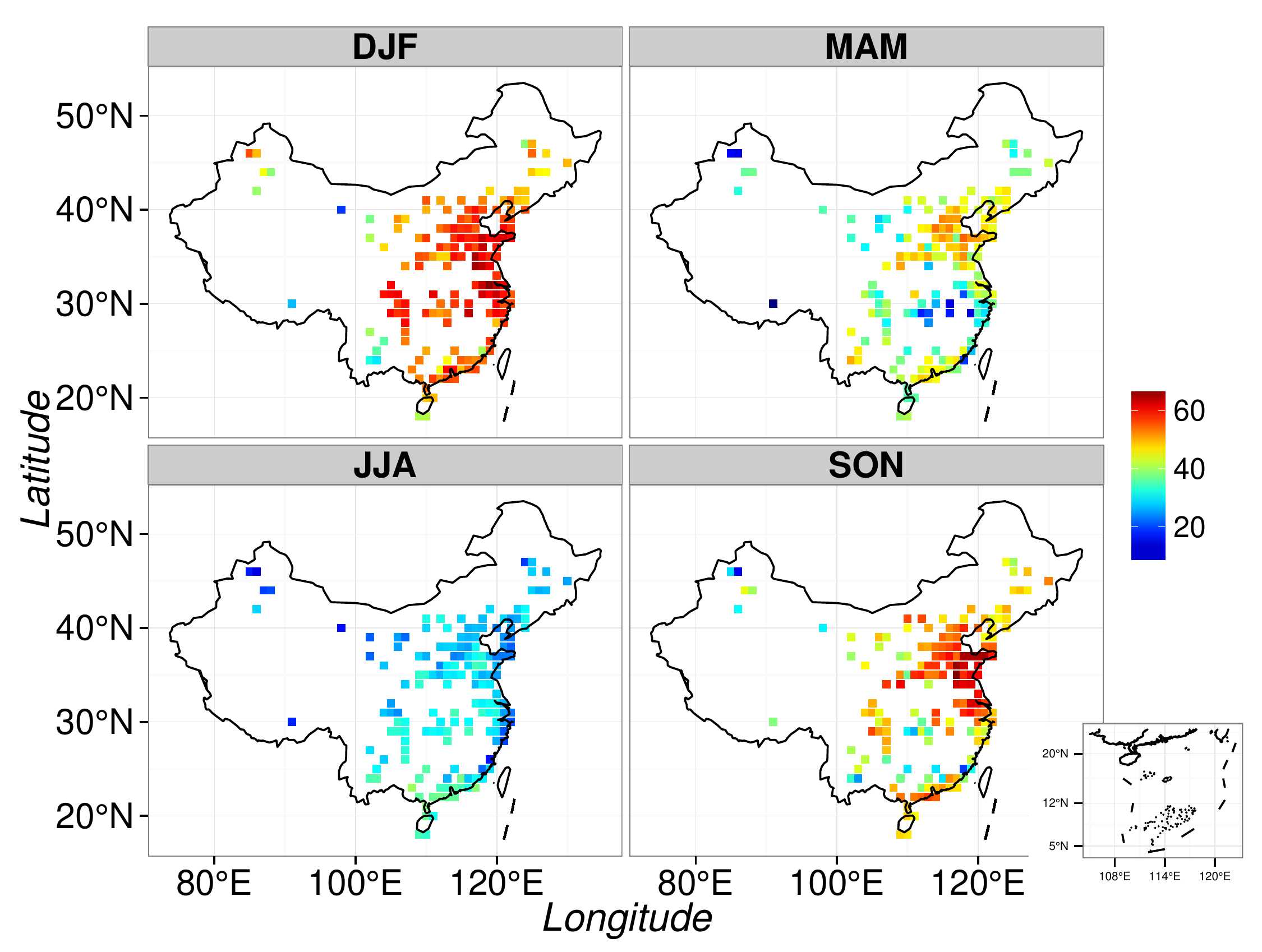}
\centering
\caption{(Color online) Distribution of weight degree in network of positive correlations. }
\label{degreeP}
\end{figure}

The weighted degree of a site, which characterizes its total correlation with surrounds, can be calculated using Eq. (\ref{Deg}). The distribution of weighted degree for positive correlations are shown in Fig. \ref{degreeP}. In comparison with Fig. \ref{mean} of mean $PM_{2.5}$ concentration, the relevance of weighted degree to mean $PM_{2.5}$ concentration can be found. In the regions with larger mean $PM_{2.5}$ concentration, the sites there have also larger weighted degree. There are the largest weighted degrees in winter as the mean $PM_{2.5}$ concentration.  

\begin{figure}
\onefigure[width=20pc]{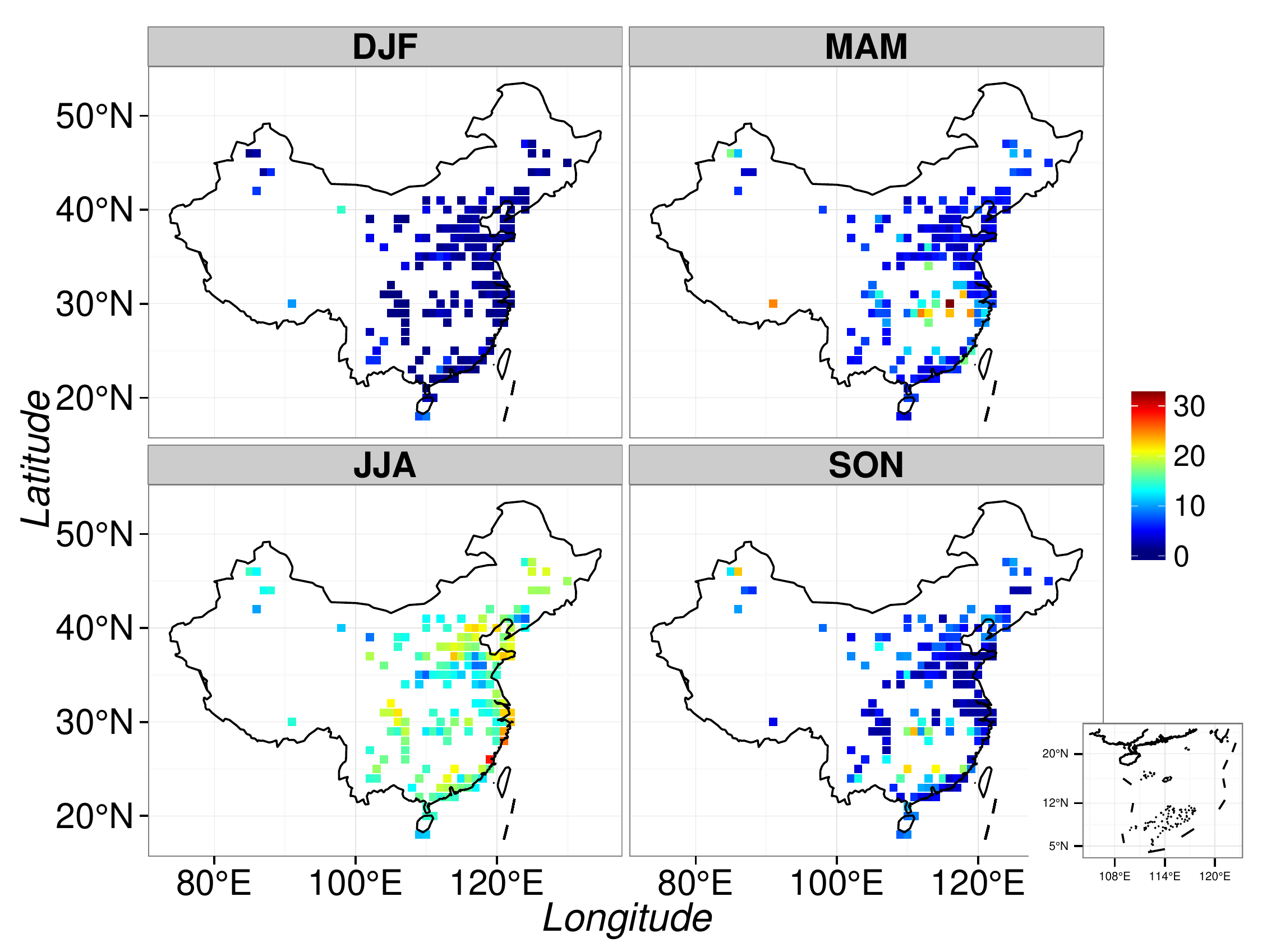}
\centering
\caption{(Color online) Distribution of weight degree in network of negative correlations.}
\label{degreeN}
\end{figure}

For negative correlations, distributions of weighted degree in different seasons are shown in Fig. \ref{degreeN}.  On the contrary, there are the largest weighted degrees in summer and the smallest weighted degrees in winter. 

\begin{figure}
\onefigure[width=20pc]{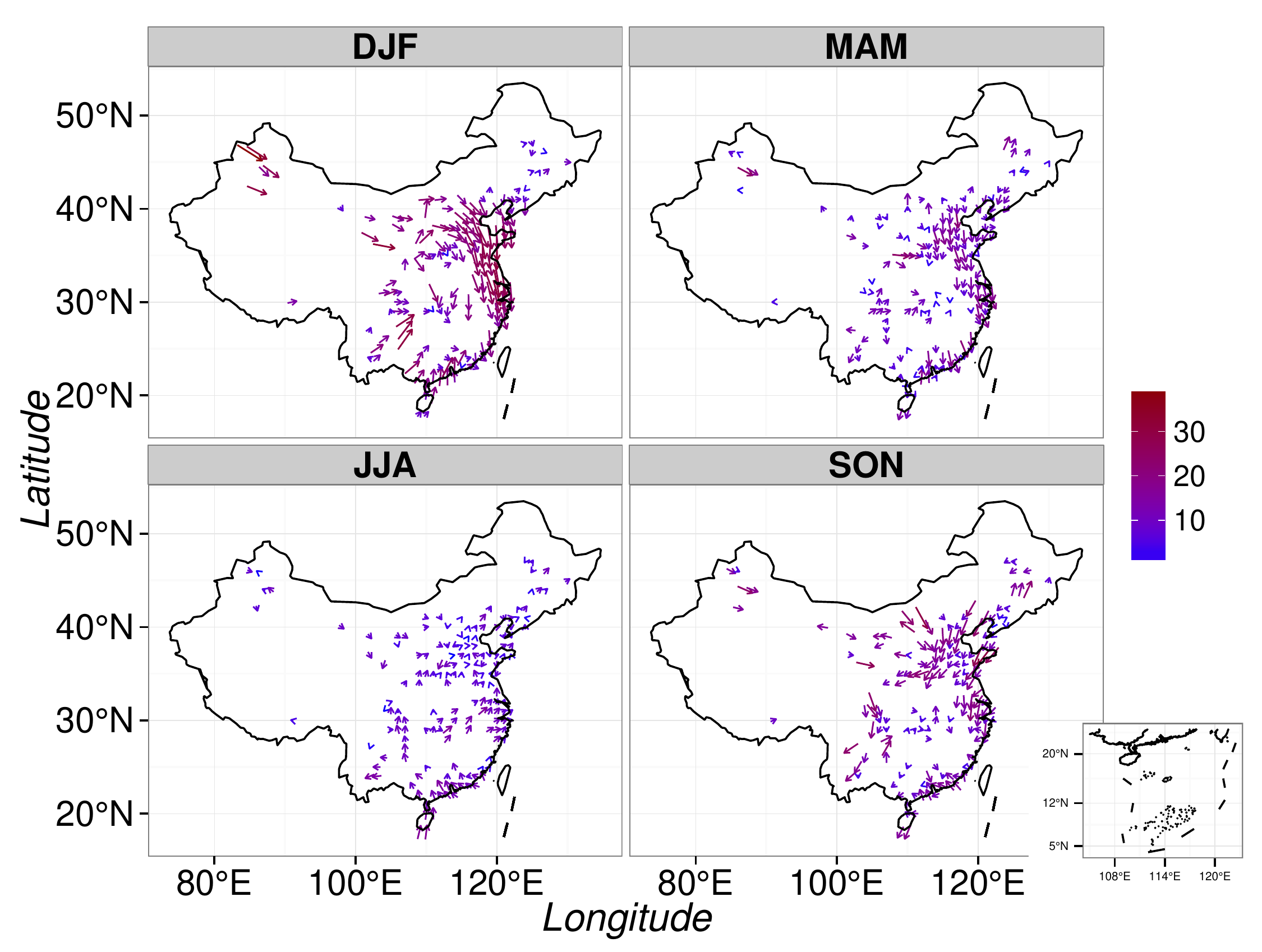}
\centering
\caption{(Color online) Distribution of directional degree in network of positive correlations.}
\label{degreeVP}
\end{figure} 

\begin{figure}
\onefigure[width=20pc]{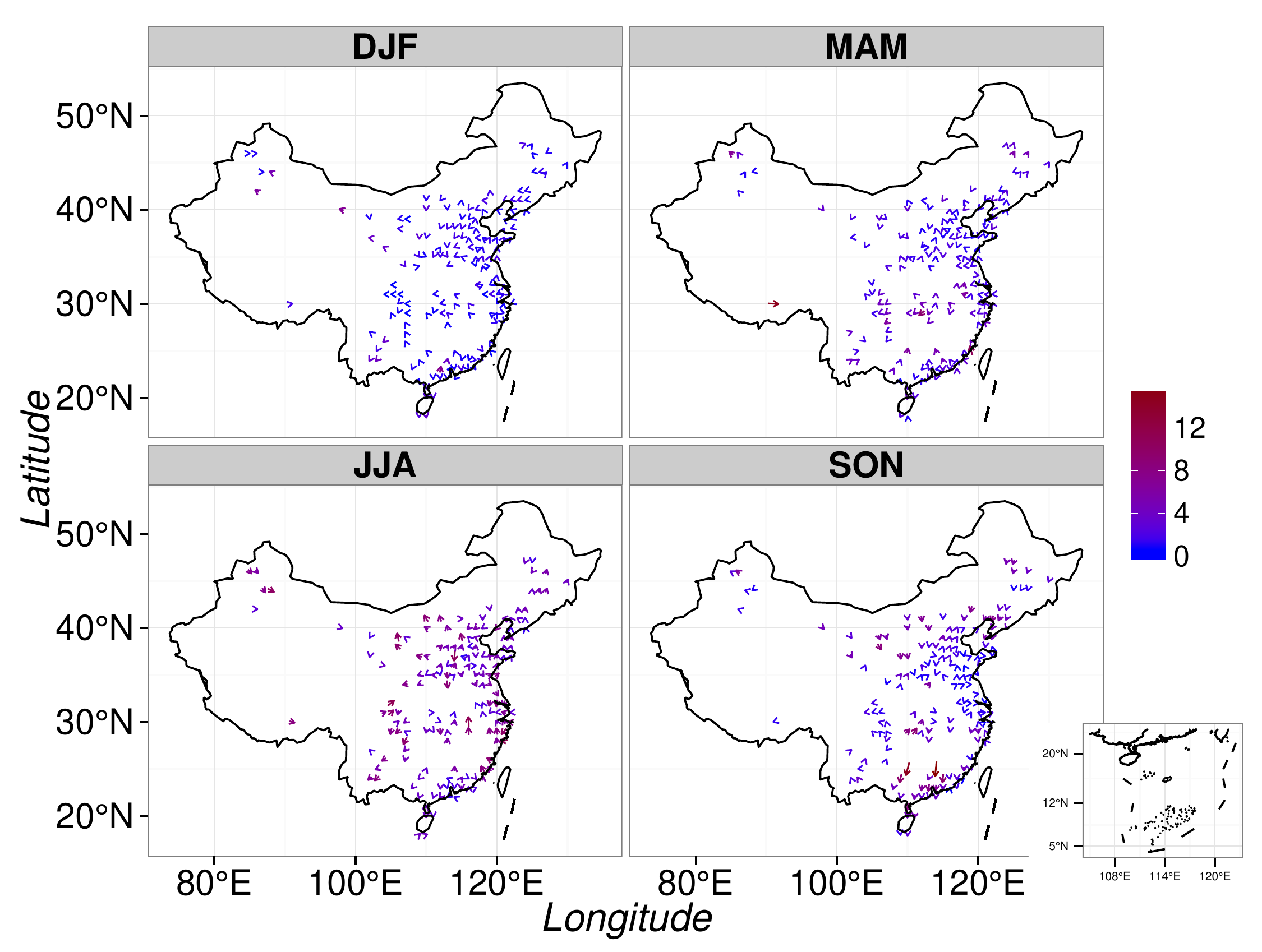}
\centering
\caption{(Color online) Distribution of directional degree in network of negative correlations.}
\label{degreeVN}
\end{figure}    

The directional degree of a site,  which is calculated according to Eq. (\ref{Vec}), characterizes its net influence to surroundings.  We present the distribution of directional degree for positive correlations in Fig. \ref{degreeVP}.  In winter, there are the strongest directional degrees in the most sites. The sites of the north-west China, such as Xinjiang, Sichuan and Guizhou, have directional degrees in the direction from west to east. The directional degrees indicate net influences of $PM_{2.5}$ concentration along Gobi and Inner Mongolia plateau, the North China Plain,  Central China , and Yangtze River Delta. This phenomena can be related to the the east Asia winter monsoon \cite{Lau,Li}, which has been shown by numerous studies. In other seasons, the directional degrees are smaller and less directional than in winter. In summer especially, only the sites around Pearl River Delta have visible directional degrees in the direction from south to north. The distribution of directional degree for negative correlations are shown in Fig. \ref{degreeVN}. No significant directional influence can be found for negative correlations.

\begin{figure}
\onefigure[width=19pc]{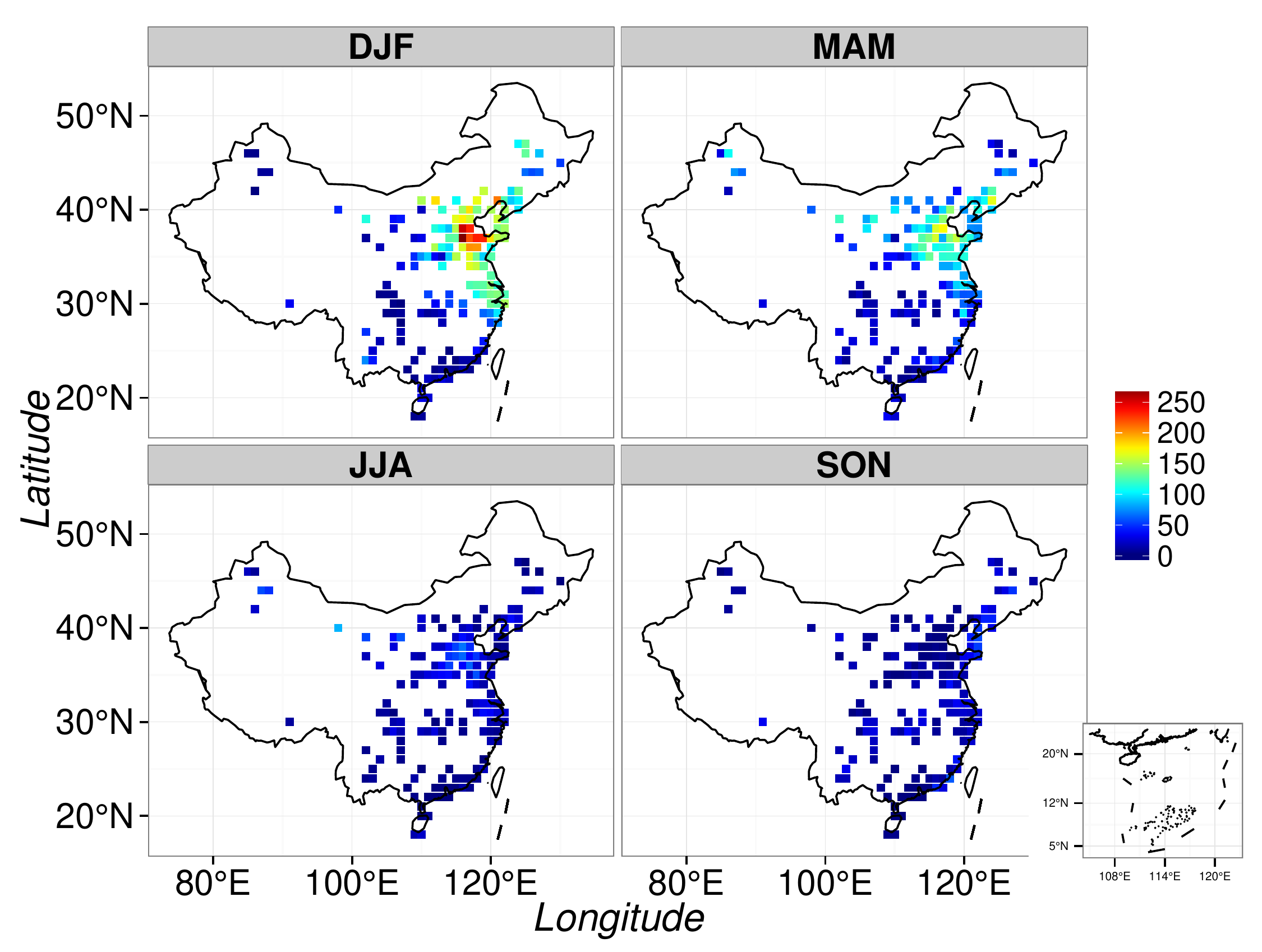}
\centering
\centering
\caption{(Color online) Distribution of weight degree in $G$ network.}
\label{degreeG}
\end{figure}

\begin{figure}[H]
\onefigure[width=19pc]{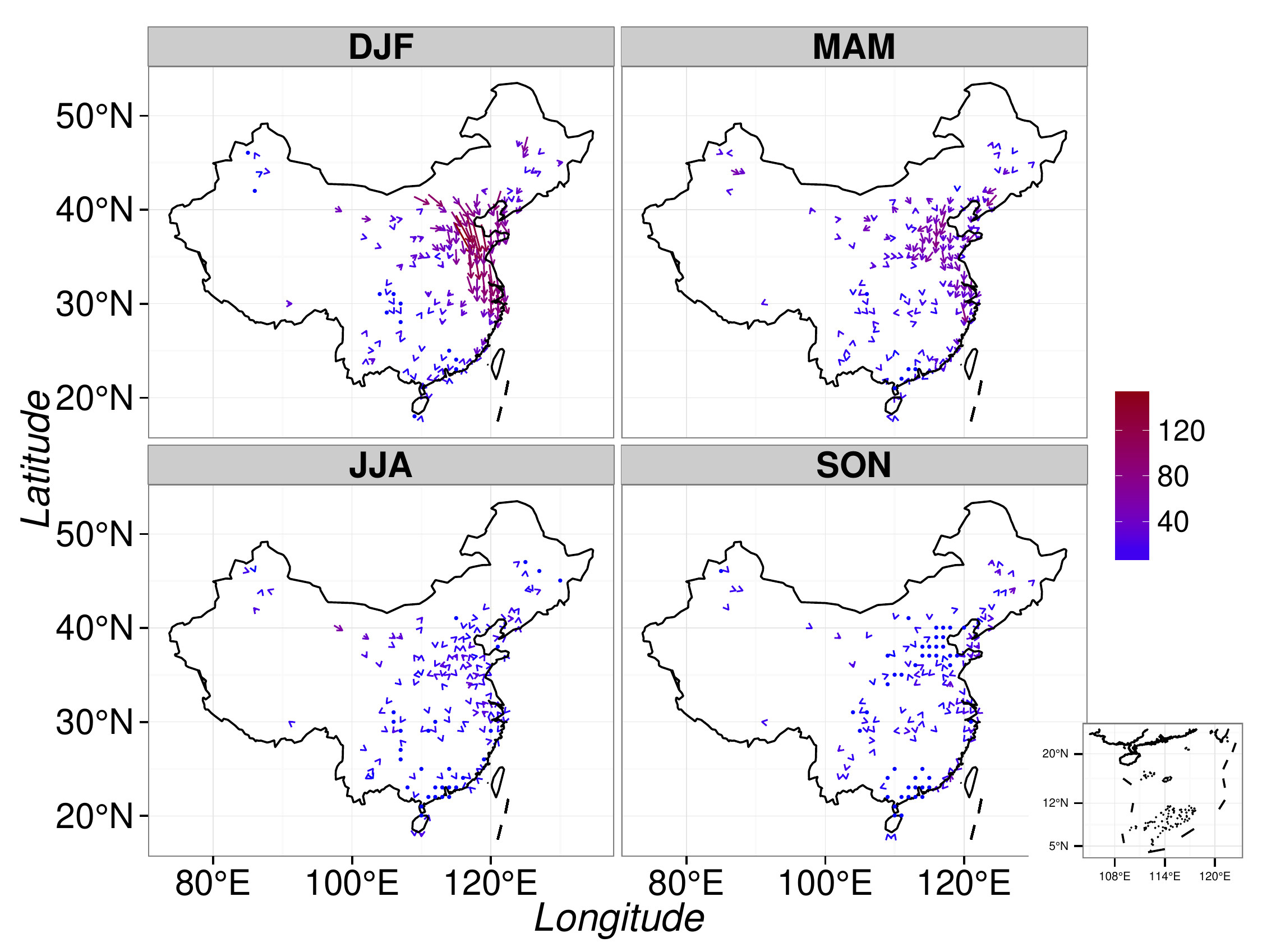}
\centering
\caption{(Color online) Distribution of directional degree in $G$ network.}
\label{vectorG}
\end{figure}

According to Eq. \ref{G}, $G_{ij}$ between sites $i$ and $j$ can be calculated. The threshold $\Theta=3.25$ of $G$ can be obtained by averaging absolute values of the shuffled data of $G$. It is found that $\Theta$ is larger than all absolute values of negative $G_{ij}$. Therefore, only a network of positive $G_{ij}$ can be defined by the adjacency matrix $B$ of Eq. \ref{admatrixB} . The distribution of weighted degree in this $G$ network is shown in Fig. \ref{degreeG}. The weighted degrees in summer and autumn are nearly zero. In winter and spring, there are large weighted degrees in the eastern part of China, especially around Beijing. The weighted degrees of $G$ network demonstrate different properties from that of $C$ network, which is shown in Fig. \ref{degreeP}. This is because of that  $G_{ij}$ characterizes actually the significance of the correlation $C_{ij}$ among $\hat C_{ij}(\tau)$ of different time lag $\tau$. The distribution of directional degrees of $G$ network is shown in Fig. \ref{vectorG}. We can see that there are large directional degrees only in winter and spring and in the eastern part of China, as the weighted degrees. The direction of directional degrees is from north to south. We think that the large weighted and directional degrees of $G$ network are resulted in by anthropogenic emissions in the region and the east Asia winter monsoon. With the $C$ and $G$ networks, different properties of $PM_{2.5}$ concentration in China have been characterized.

\section{Summary}  
\label{Summary}

We have studied the correlations of $PM_{2.5}$ concentrations in different sites of China. Using the hourly $PM_{2.5}$ concentration data in 754 monitoring sites over China from Dec. 2014 to Nov. 2015, we can calculate the  correlations between different sites in four seasons. The probability distribution functions of positive and negative correlations depend on season. With averages and standard deviations of correlation, the different probability distribution functions of different seasons can be scaled into one scaling function. This indicates that there is maybe the same mechanism related to correlation of $PM_{2.5}$  concentration in different seasons. The positive correlations are resulted by the transport of $PM_{2.5}$. For positive correlations, there are the largest average in winter and the smallest average in summer. The negative correlations are caused probably by large scale oscillating climate conditions. In opposite to positive correlations, there are the largest average in summer and the smallest average in winter for negative correlations.

Further, $PM_{2.5}$ concentrations in different sites of China are studied from the aspect of complex network. Networks of $PM_{2.5}$ concentration can be defined either by correlations or by their significances. From weighted and directional degrees of network, different properties of $PM_{2.5}$ concentration can be studied. In the networks of positive correlations, the largest weighted degrees appear in winter and in the North China plain as far as location is concerned. The location distribution of weighted degree and its seasonal dependence are in accord with that of mean $PM_{2.5}$ concentration. In the networks of negative correlations, the largest weighted degrees appear in summer. This indicates further that the origins of positive and negative correlations are different. Significant directional degrees are found for positive correlations in winter. They demonstrate the existence of net influences of $PM_{2.5}$ concentrations along Gobi and inner Mongolia plateau, the North China Plain, Central China, and Yangtze River Delta. From significances of positive correlation, we can define a network which has large weighted and directional degrees only in winter and spring and in the eastern part of China. The directional degrees are in the direction from north to south. These properties of $PM_{2.5}$ concentrations could be related to anthropogenic emissions in the region and the Asia winter monsoon.

\acknowledgments
We are grateful to the financial support by Key Research Program of Frontier Sciences, CAS (Grant No. QYZD-SSW-SYS019) and the fellowship program of the Planning and Budgeting Committee of the Council for Higher Education of Israel, the  Israel Ministry of Science and Technology (MOST) with the Italy Ministry of Foreign Affairs, MOST with the Japan Science and Technology Agency, the BSF-NSF foundation, the Israel Science Foundation, ONR and DTRA. Yongwen Zhang thanks the postdoctoral fellowship funded by the Kunming University of Sciences and Technology and Dean CHEN thanks the China scholarship council. We also acknowledge the data resources provided by the ministry of environmental protection of China(http://113.108.142.147:20035/emcpublish/).


\end{document}